\documentclass[authoryear,preprint,review,12pt]{elsarticle}
\usepackage{amssymb,amsmath,lineno}
\usepackage[margin=3cm]{geometry}

\journal{Geochimica et Cosmochimica Acta}

\newcommand{\bv}[1]{\mbox{{\bf #1}}}

\newcommand{\ud}{\textrm{d}}

\newcommand{\cmyr}{\mathrm{cm/yr}}

\newcommand{\DTCO}{\Delta T_C}
\newcommand{\Uo}{U_0}

\newcommand{\gravity}{\bv{g}}

\newcommand{\DTC}{\Delta T_C}

\newcommand{\msp}{\langle v_p \rangle}
\newcommand{\degC}{^{\circ}{\textrm{C}}}

\newcommand{\drho}{\Delta \rho}

\newcommand{\vf}{\bv{v}_f}
\newcommand{\vm}{\bv{v}_m}

\newcommand{\Cm}{C_m}
\newcommand{\Co}{C_0}

\newcommand{\Cref}{C_0}

\newcommand{\Th}{^{230}\mathrm{Th}}
\newcommand{\Ra}{^{226}\mathrm{Ra}}
\newcommand{\Pb}{^{210}\mathrm{Pb}}
\newcommand{\Pa}{^{231}\mathrm{Pa}}

\begin{document}
\begin{frontmatter}

  \title{Melt transport rates in heterogeneous mantle beneath
    mid-ocean ridges}

  \author[Oxford]{Samuel M. Weatherley\corref{cor1} \fnref{GEUS}}
  \ead{smw@geus.dk} \cortext[cor1]{Corresponding author}
  \author[Oxford]{Richard F. Katz} \ead{richard.katz@earth.ox.ac.uk}
  \address[Oxford]{Department of Earth Sciences, University of Oxford,
    South Parks Road, Oxford. OX1 3AN. UK} \fntext[GEUS]{Present
    address: GEUS, Geological Survey of Denmark and Greenland,
    {\O}ster Voldgade 10, 1315 Copenhagen K. Denmark. Tel: +45 91 33
    38 68}

\begin{abstract}
  Recent insights to melt migration beneath ridges suggest that
  channelized flow is a consequence of melting of a heterogeneous
  mantle, and that spreading rate modulates the dynamics of the
  localized flow.  A corollary of this finding is that both mantle
  heterogeneity and spreading rate have implications for the speed and
  time scale of melt migration.  Here, we investigate these
  implications using numerical models of magma flow in heterogeneous
  mantle beneath spreading plates.  The models predict that a broad
  distribution of magma flow speeds is characteristic of the sub-ridge
  mantle. Within the melting region, magmatic flow is fastest in
  regions of average fusibility; surprisingly, magmas from sources of
  above-average fusibility travel to the ridge in a longer
  time.  Spreading rate has comparatively simple consequences, mainly
  resulting in faster segregation speeds at higher spreading rates.
  The computed time scales are short enough to preserve deep origin
  $\Th$ disequilibria and, under favourable parameter regimes, also
  $\Ra$ excesses.  An important prediction from the models is that
  mantle heterogeneity induces significant natural variability into
  U-series disequilibria, complicating the identification of
  relationships between disequilibria and ridge properties or chemical
  signatures of heterogeneity.
  
\end{abstract}

\end{frontmatter}

\setcounter{figure}{0} \renewcommand{\thefigure}{\arabic{figure}}

\section{Introduction}

Chemical heterogeneity in Earth's mantle induces local and regional perturbations in the fusibility and melting
rate beneath oceanic volcanoes.  For example, regions
enriched in subducted oceanic lithosphere start to melt at deeper depths and with
higher melt productivity than mantle peridotite
\citep{Hirschmann1996}, whereas other regions are so refractory, and
contribute so little to the genesis of magmas, that they are
essentially unsampled by mid-ocean ridge basalts
\citep{Liu2008}. In two recent numerical studies \citet{Weatherley2012}
and \citet{Katz2012} considered the effect that mantle heterogeneity
has on the dynamics of melt migration beneath oceanic volcanoes.
Their models suggest that (i) heterogeneity causes magma to localize
into a network of high porosity channels, and (ii) that the topology
of the channel network depends on the spatial arrangement of mantle
heterogeneity.   Channelised flow is often invoked to explain the
major and trace element characteristics of oceanic basalts and mantle
rocks \citep[e.g.][]{Kelemen1995, Kelemen1997}, and to explain
geochemical evidence for rapid magma flow speeds in the mantle
\citep[see reviews by][]{Lundstrom2005, Elliott2014}.  In this paper
we explore the prediction arising from \citet{Weatherley2012}
and  \citet{Katz2012} that mantle heterogeneity has important but as yet
unrecognized consequences for the time scales of melt migration beneath
oceanic volcanoes. 

Temporal constraints on magma flow in the mantle largely are derived
from observations of U-series disequilibria in very young or zero age
MORB.  Specifically, disequilibria provide temporal constraints on
combined melt generation and transport processes that have occurred
within approximately 5 half-lives of the shorted-lived nuclide in a
parent--daughter nuclide pair.  For
understanding magma flow processes, the most useful nuclides are
$\Th$, $\Pa$, $\Ra$ and $\Pb$; these decay with half-lives of
c.~75~ka, 33~ka, 1.6~ka and 22 yr.  Since a wide range of processes can
generate U--series disequilbria, models are often used to explore
the implications of disequilibria for the time scale of melt migration
in the mantle.  In a review of existing attempts to model U-series melting models,
\citet{Elliott2014} note that models generally require a large set
of assumptions such as melt-mineral partition coefficients, mantle
porosity, source mineralogy and petrology, the way in which melt is
organized in the mantle, and the location and mechanism of chemical
fractionation.  Unsurprisingly, the combined outcome of these models is rather complex.
Interpretations of $\Th$ disequilibria suggest that the
mean flow speed between the melt source and ridge axis is around
1~m/yr, and shorter lived $\Ra$ disequilibria can be interpreted to
require speeds $>$50~m/yr \citep{Rubin1988, Kelemen1997, McKenzie2000,
  Stracke2006, Turner2010}. These speeds are fast enough to require
magma flow by some form of channelised melt transport. However,
existing models of chanelized flow appear to be incapable of
reproducing the $\Pb$ disequilibria signatures observed by
\citet{Rubin2009}, thus $\Pb$ deficits are unexplained by
current theories of channelized melt migration.

How U-series disequilibria in MORB relate to chemical signatures of
mantle heterogeneity is the focus of only a small number of
investigations.  In a study of the Southeast Indian Ridge,
\citet{Russo2009} investigated nonsystematic geographic variation of
U-series disequilibria with axial depth, crustal thickness and ridge
morphology. The authors explain the observations by along-axis
variation in the melt supply, which they suggest is a response to long
wavelength temperature gradients and small variations in source
heterogeneity.  Several other studies note a weak, positive
correlation in noisy data between the amplitude of $\Th$ disequilibria
and enrichment of the source in incompatible elements
\citep{Lundstrom1995, Lundstrom2000, Elliott2003, Kokfelt2003,
  Koornneef2012}.  Since Th and U are fractionated more strongly in
the presence of garnet, one explanation for this trend is that larger
amplitude disequilibria result from a relatively larger proportion of
melt originating in the garnet stability field
\citep{Lundstrom1998}. Yet, as \citet{Russo2009} demonstrate, it is
non-trivial to make an explicit connection between the amplitude of
disequilibria and magmas from garnet-bearing source rocks.
Furthermore, it is unclear how the dynamics and time scale of magma
flow might contribute to the observed signals.

The current dynamical understanding of melt migration is founded on
mathematical theories and numerical models for the physics of magma
flow in porous, compactible mantle rock \citep{McKenzie1984,
  Fowler1985, Ribe1985, Bercovici2001}.  A goal for many models of
magma flow is to explore and understand the dynamics of melt
localization. Among these, channel formation is attributed to the
reaction infiltration instability \citep{Ortoleva1987} that in the
context of melt migration, describes a positive feedback between magma
flux and dissolution of porous mantle rock.  This feedback causes a
net transfer of mass from the solid mantle to the magma, locally
increasing the porosity and melting rate, and results in the formation
of channels \citep{Kelemen1990, Kelemen1995, Aharonov1995}.  As a
consequence, models of this process must treat the physics and
thermochemistry of magma flow as a coupled system.  For simplicity,
previous studies generally restrict the problem to magma flow in an
upwelling column of partially melting mantle rock, neglect
conservation of energy and parameterise melting using kinetic theory
\citep{Spiegelman2001, Spiegelman2003, Liang2003, Liang2010,
  Schiemenz2011, Hesse2011}.

Other models by \citet{Katz2008, Katz2010}, \citet{Hewitt2010},
\citet{Weatherley2012} and \citet{Katz2012} differ by adopting an
energetically consistent approach to melting.  These models capture
the onset of partial melting and suggest that latent heat effects
suppress the emergence of channels in the situation of magma flow
through a chemically homogenous mantle.  However, models by
\citet{Weatherley2012} and \citet{Katz2012} show that channels emerge
in a heterogeneous mantle, where spatial variations in composition are
considered to be synonymous with variations in the fusibility of
mantle rock. \citet{Weatherley2012} shows that more fusible
heterogeneities melt preferentially, and that the resulting flux of
magma can be sufficient to nucleate high porosity channels.  In a
companion study, \citet{Katz2012} considers the effect of different
topologies of mantle heterogeneity and plate spreading on melt
migration; they find that both heterogeneity and plate spreading have
important consequences for the arrangement of channels and delivery of
melt to the ridge axis.

In this contribution we extend the models of
\citet{Katz2012} to quantify the effect of source heterogeneity and
spreading rate on the speed and time scale of melt migration.  Section
~\ref{sec:model} outlines the theory and numerical model of magma flow
in heterogeneous mantle rock, summarises the petrological system
(which is a key component of the model) and provides details on the
approach used to determine the speed and time scale of melt migration.
Solutions to the numerical experiments are presented in
section~\ref{sec:results}.  Results are presented in dimensional form.
Those presented at the start of the section aim to characterise the
time scales and flow speeds predicted by the models, while those at
the end of the section illustrate the effect of heterogeneity.  The
results are discussed in section~\ref{sec:discussion}, where
particular attention is paid to how the time scale of melt migration
depends on spreading rate and mantle heterogeneity.  A scaling
analysis explores how the results might change under different
parameter regimes; the assumptions and limitations inherent in the
models are also noted.  The section concludes with a discussion on the
implications of the results for our understanding of U-series
disequilibria in MORB, including tentative prediction of their
covariance with source composition.

\section{The model}
\label{sec:model}

\subsection{Model configuration}

\begin{figure*}
  \centering
  \includegraphics{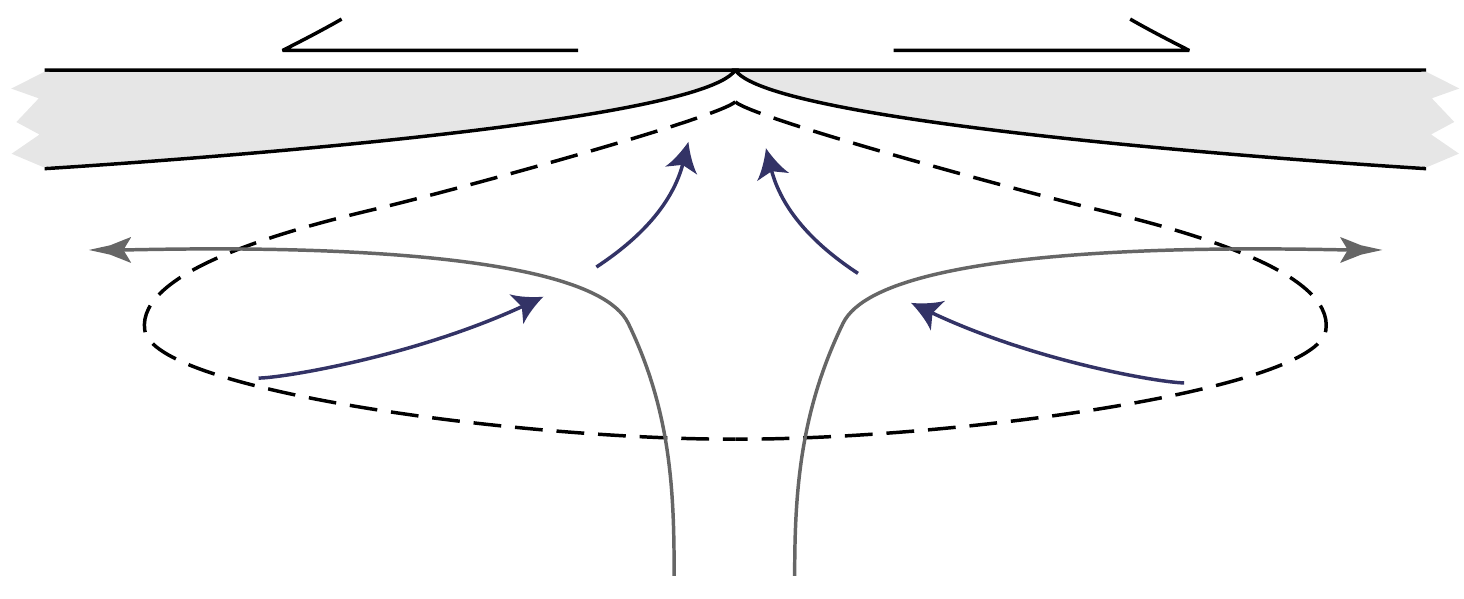}
  \caption{A schematic illustration of the model setup.  Lithospheric
    plates (light grey regions) spread apart at a known rate, driving
    large scale flow in the solid mantle (dark grey arrows).
    Decompression causes the mantle to melt within an unknown region
    indicated by the dashed line.  Within this region solid rock and
    magma coexist and the melt segregates by porous flow (blue
    arrows)}
  \label{fig:setup}
\end{figure*}

The situation we consider is of magma flow, melting and melt--rock
interaction in passively flowing, compositionally heterogeneous mantle
beneath mid-ocean ridges (figure~\ref{fig:setup}).  Subsolidus mantle
rock flows into the bottom of the domain, upwells and melts
adiabatically, and the melt then makes its way towards the surface
where it is erupted or freezes into the lithosphere.  Previous models
by \citet{Katz2008, Katz2010} and \citet{Katz2012} form the basis for
those presented here, although they are modified in order to derive
the predictions of the speed and time scale of melt migration beneath
ridges. A brief summary of the behaviour predicted by the model is
given in section \ref{sec:behaviour}.

The models solve equations for conservation of mass, momentum,
composition and energy in a thermally and chemically diffusive system
with two phases and two thermodynamic components.  In other words,
they seek solutions for the coupled dynamics of solid mantle rock and
liquid magma, and employ a basic petrological model to allow for
melting and freezing. In the model framework mantle rock is considered
to be a compactible, crystalline solid that deforms by creeping flow;
magma is modelled as a low viscosity liquid that migrates by porous
flow.  In partially molten regions, the magma and mantle are assumed
to coexist and interpenetrate \citep{McKenzie1985, Fowler1985,
  Ribe1985, Bercovici2001}. The boundary conditions for the models are
identical to those used by \citet{Katz2010}, the only difference being
that the internal boundary condition for melt extraction is pinned to
the apex of the melting region, rather than to the boundary at the
mid-line of the domain.  Full details for this condition are provided
in \citet{Weatherley2013}.  An additional deviation between the models
here and previous variants is that mantle flow is considered to be a
passive response to plate spreading; variation in the gravitational
body force on the two-phase aggregate is neglected.  The equations are
discretized using a finite volume approach on a staggered mesh with
1.25~km grid spacing in each direction. The governing equations are
solved using the method described by \citet{Katz2012}.

\begin{figure*}
  \centering
  \includegraphics{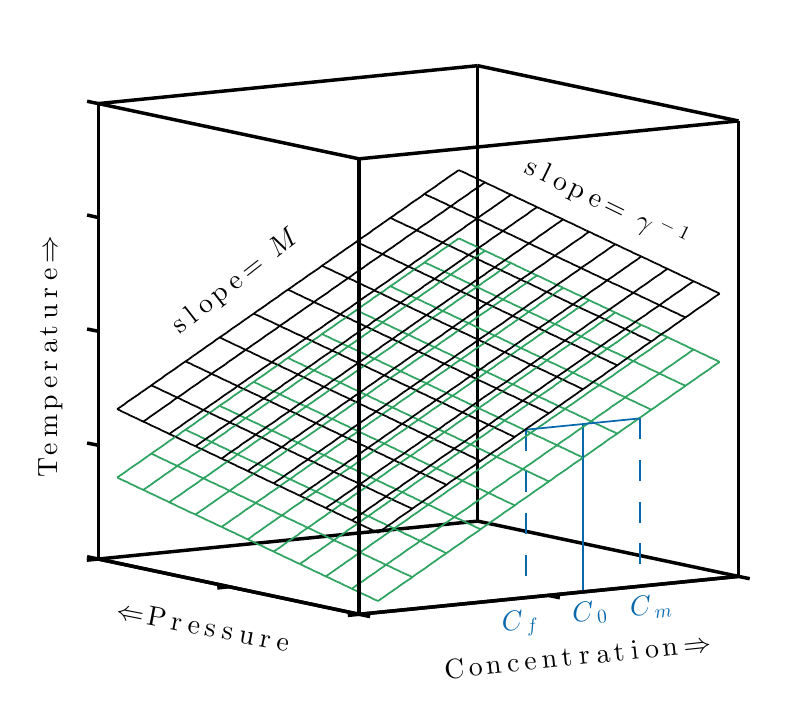}
  \caption{A binary solid solution with linearized solidus (lower surface, green
    lines) and liquidus (upper surface, black lines) constitutes the
    phase diagram.  $C_0$ marks the bulk composition at zero pressure,
    $C_f$ and $C_m$ indicate the compositions of coexisting magma and
    rockm also at zero pressure.  The mass fraction of the liquid is
    $(C_0 - C_f)/(C_m - C_f)$.}
  \label{fig:phase}
\end{figure*}

\subsection{Phase relations and mantle heterogeneity}
\label{sec:pet}

Mass transfer between the solid and liquid by melting and freezing is
an important feature of the model.  To allow for this, we follow the
examples of \citet{Ribe1985, Katz2008, Katz2010, Hewitt2010, Katz2012} and
\citet{Weatherley2012} and assume that coexisting rock and magma are
in local thermodynamic equilibrium. An important consequence of this
assumption is that magma and mantle rock are in local chemical and
thermal equilibrium in all partially molten regions.  A simple phase
diagram (figure~\ref{fig:phase}) comprising two components of
different fusibility describes the petrological system.  At every time
step the phase diagram is applied in each grid cell in order to relate
the bulk composition, bulk enthalpy, and lithostatic pressure to the
temperature, phase compositions, and rate of mass transfer between
rock and magma.  Since the phase diagram corresponds only to a
hypothetical, highly simplified petrologic system, compositional
variations are expressed in terms of their perturbation to the solidus
temperature relative to a reference composition, or equivalently referred
to here as fusibility. This is expressed as $\DTC = M (\Cm - \Co)$, where $M$
is the slope $\partial T / \partial C$ of the solidus, $\Cm$ is the
local composition of the mantle rock and $\Co$ is a reference
composition.

\begin{figure*}
  \centering
  \includegraphics{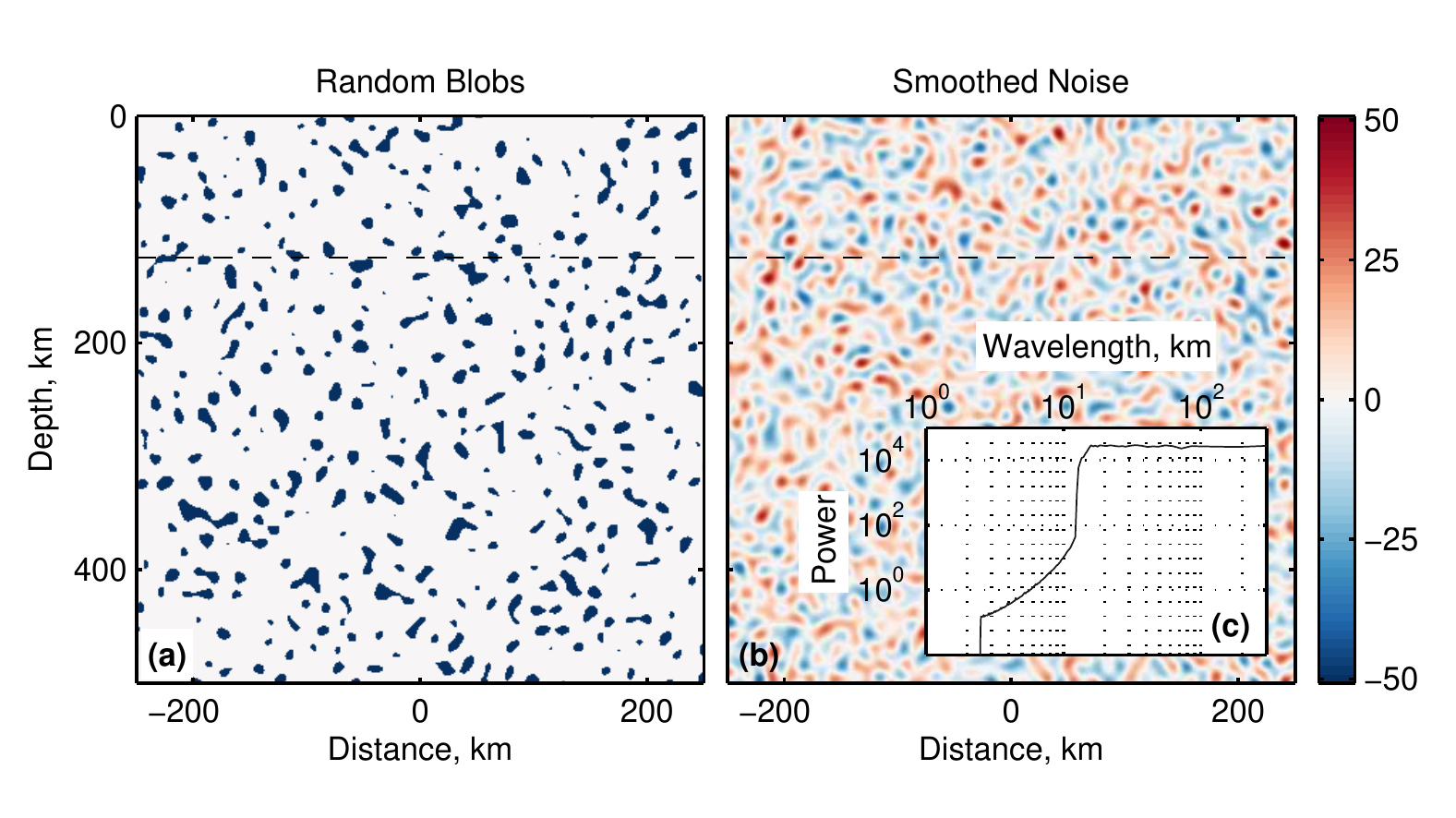}
  \caption{The compositional heterogeneity fields used to initialize
    the models, shown as a perturbation to the solidus temperature,
    $\DTC = M_S (\Cm - \Co)$.  $C_0 = 0.5$ for both fields.  Dashed
    lines in (a) and (b) show the size of the computational domain
    relative to the heterogeneity fields.  The heterogeneity fields
    are larger in the $z$--direction in order to evolve the bottom
    boundary condition on $C$ as material is advected into the domain.
    (a) `Random blob' style heterogeneity .  $10\%$ of the area is
    occupied by more fusible material.  (b) `Smoothed noise' type
    heterogeneity. (c) Directionally averaged power spectrum for the
    heterogeneity field in panel (b).  It shows approximately uniform
    power for wavelengths $>11$ km and greatly reduced power for
    shorter wavelengths. }
  \label{fig:het}
\end{figure*}

Figures \ref{fig:het}a and \ref{fig:het}b show the two contrasting
heterogeneity fields used to initialize the models.  They are
constructed using stochastic algorithms that generate spatial
variations in the ratio of the two thermodynamic components.  Full
details of the algorithms are provided in \citet{Katz2012}. In
figure~\ref{fig:het}a, the `random blob' heterogeneity model consists
of randomly shaped, isolated regions that are uniformly enriched in
the more fusible component.  Prior to the onset of melting, the
enriched regions perturb the local solidus temperature by $-$50~K,
have a characteristic size of 10~km and occupy an area fraction of
0.1. The composition of the non-enriched matrix defines $\Co$.  The 
`smoothed noise' model (figure~\ref{fig:het}b) is produced by allowing
the ratio of the components to vary smoothly and continuously over all
wavelengths greater than a specified cut-off.  The spectrum has equal
power at all wavelengths greater than 10~km and negligible power for
wavelengths $<$5 km (figure \ref{fig:het}c).  Prior to the onset of
melting, $\DTC$ varies by up to  $\pm$50~K and $\Co$ is set to the 2-norm of
the heterogeneity field.  The inspiration for these conjectural
descriptions of mantle heterogeneity lies in (a) recognition that
regions of the mantle enriched in ancient oceanic crust are more
fusible than those that are not \citep{Hirschmann1996}, and that the
mantle contains refractory domains that are essentially unsampled by
melting beneath oceanic volcanoes \citep{Liu2008}, (b) uncertainty
surrounding the lithology of mantle heterogeneity, and whether
recycled oceanic crust melts and enriches the surrounding mantle prior
to the main melting event beneath ridges \citep[e.g.][]{Yaxley1998}, (c)
uncertainty in the topology of heterogeneity 
\citep{Allegre1986, Kellogg1990}, (d) analyses of scattered seismic
waves that suggest chemically distinct heterogeneities with length
scales of around 10 km exist in the mantle \citep{Kaneshima1999,
  Kaneshima2003}, (e) geochemical estimates that the MORB source
region contains around 10$\%$ recycled oceanic crust
\citep[e.g.][]{Wood1979, Sobolev2008, Shorttle2011}, and (f) the need
to represent mantle heterogeneity on scales larger than the grid
resolution of the simulations (1.25 km).


\subsection{Determination of melt migration timescale}

To determine the speed and duration of melt migration predicted by the
model, the domain is seeded with Lagrangian particles that track the
motion of individual fluid parcels.  The magma velocity is given by
\begin{equation}
  {\bf{v}}_f = {\bf{v}}_m - \frac{K}{\phi \mu} \left[ \nabla
    P_d + \Delta \rho \gravity \right],
  \label{eq:time-vf}
\end{equation}
where $K=k_0\phi^n$ is permeability ($k_0$ is the permeability
coefficient, $\phi$ is porosity and $n$ is the permeability exponent)
$\mu$ is the viscosity of the magma, $P_d$ is the dynamic fluid
pressure, $\Delta \rho$ is the density difference between the solid
mantle and liquid magma and $g$ is the acceleration due to
gravity. Equation~\ref{eq:time-vf} states that the velocity of 
the magma depends on the matrix velocity, dynamic pressure gradients,
and buoyancy forces that are modulated by permeability and the
viscosity of the magma.  The tracer particles are defined to be
perfectly incompatible; they are released from the solid at the onset
of melting, can be re-incorporated into the solid by freezing, and can
exit the domain at the ridge axis.  The particles are accurately
tracked by updating their positions at each time step in a two stage
process. First the velocity field is linearly interpolated to the
position of each particle. Then using the interpolated velocity, the
position of each particle is updated over one half of a time step.
The process is then repeated and the tracer particles are moved over
the remaining half of the time step.

The tracer particles are initialized by placing one in the centre of
every grid cell.  As the particles move through the domain, they record 
at every time step their velocity, position, the porosity and
composition of the matrix within their host grid cell, and the model
time.  Those that exit the domain at the ridge axis are filtered to
exclude any that ({\textit{i}}) are not initialized in subsolidus
mantle for which the degree of melting is zero, and ({\textit{ii}})
are refrozen into the matrix at any point between the melt source and
the ridge axis.  Data from particles that pass through this filter are
used to compute the duration and speed of melt migration. In
subsolidus regions, the motion of the tracer particles is equal to the
solid mantle flow field.  Since the flow field approximates the corner
flow solution, the upwelling rate is fastest directly beneath the
ridge axis and slower in the flanking regions \citep{Batchelor1967}.
On long time scales a consequence of this is that the rate of delivery
of tracer particles  to the melting region is uneven across its width.
To reduce any sampling bias the tracer particles are reinitialized periodically on a
time scale of 350 kyr.  Particles are not reinitialized more
frequently, since magma migrating to the ridge on shorter time scales
are capable of preserving $\Th$ disequilibria generated at the base of
the melting region.

\section{Results}
\label{sec:results}

\subsection{Overview of dynamics}
\label{sec:behaviour}

This study is based around three suites of numerical experiments.  One
suite is initialized with a chemically homogenous mantle, another with
the random blob heteroegenity model (figure~\ref{fig:het}a), and the
final suite with the smoothed-noise heterogeneity
model (figure \ref{fig:het}b).  Each suite comprises four
simulations with half-spreading rates $U_0$ of 2, 4, 6, and 8~cm/yr.
Some key aspects of the model behaviour are outlined below, but the
reader is referred to \citet{Katz2012} for a complementary description and
analysis.

\begin{figure*}
  \centering
  \includegraphics{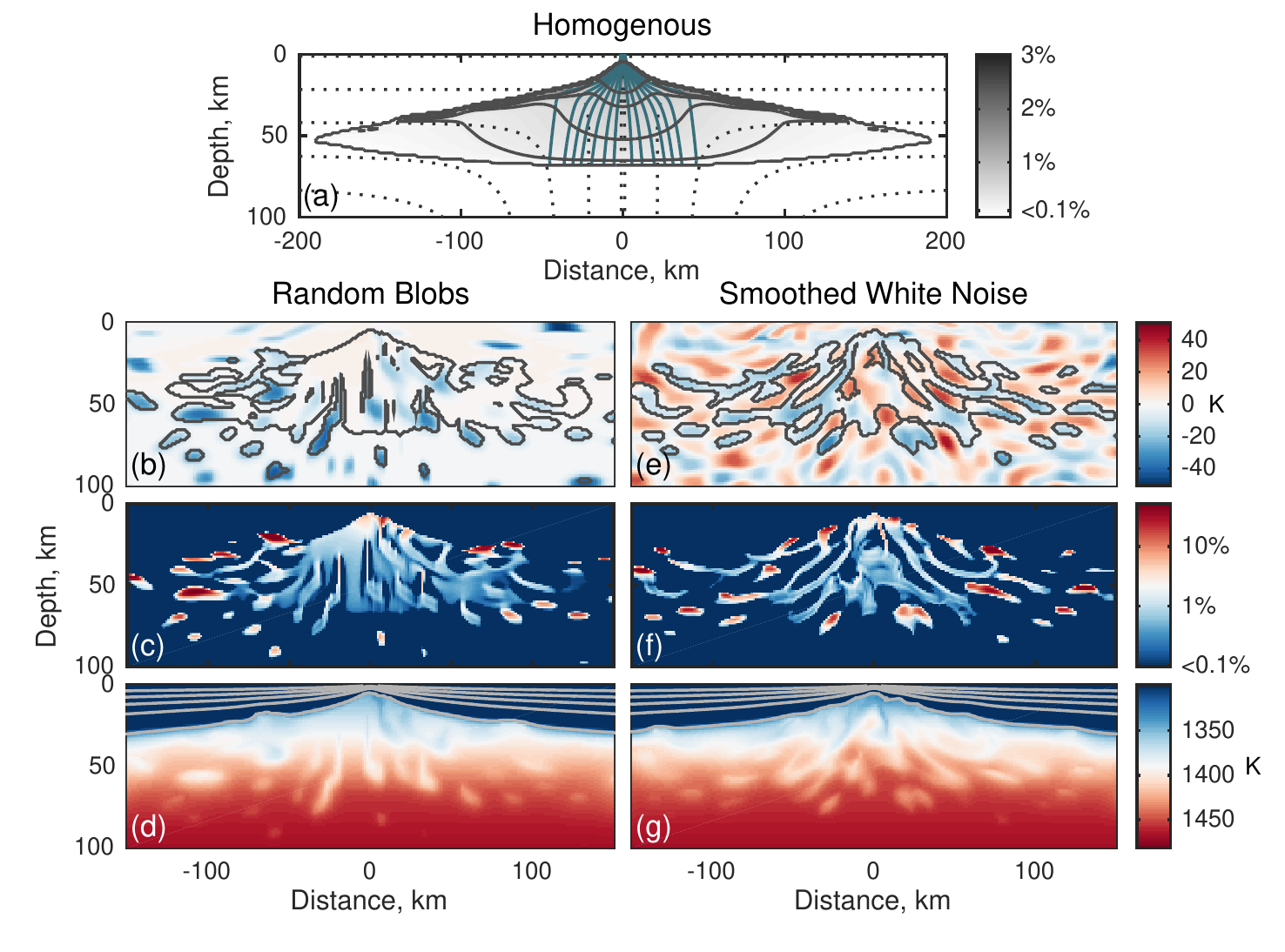}
  \caption{Snapshots from representative simulations run with
    contrasting styles of mantle heterogeneity.  In each case the half
    spreading rate is 4 cm/yr and the model time is 1.9 Ma. Panel (a)
    shows results from a simulation run with a homogenous mantle.
    Colours and solid dark grey contour lines show the porosity and
    extent of the melting region; the dotted streamlines show the
    solid mantle flow field; the solid blue-green lines plot the
    trajectories of some tracer particles that travel to the ridge
    axis.  Panels (b)--(d) show results from a simulation with the
    random blob type heterogeneity, and panels (e)--(g) show results
    from a simulation run with the smoothed noise heterogeneity.
    Panels (b) and (e) express mantle heterogeneity as a perturbation
    to the solidus (grey line) temperatures, (c) and (f) show
    porosity, and d and g show temperature.  The solid grey lines
    contour temperatures between 0$\degC$ and 1300$\degC$.  The models
    are subject to the parameter regime outlined in table
    \ref{tab:ridge-p-vals}. }
  \label{fig:ridge}
\end{figure*}

Figure \ref{fig:ridge}a illustrates the basic dynamical behaviour of
the system with a snapshot from a simulation run with no initial
heterogeneity and a half spreading rate of 4 cm/yr.  Models run with a
homogenous mantle do not predict channelized flow; instead melt
migrates by dynamically stable, diffuse porous flow.  Melting is
confined to a roughly triangular-shaped region extending about 75~km
deep into the mantle and over 100~km to either side of the ridge
axis. Throughout this region, coexisting magma and mantle rock are in
local chemical and thermodynamic equilibrium (section \ref{sec:pet}).
Compaction maintains the porosity at low values, typically around
$<1\%$. However, the porosity varies systematically throughout the
melting region, largely as a response to the pattern of mantle
upwelling; porosity is largest directly beneath the ridge axis where
the rate of mantle upwelling (and decompression) is greatest, and
decreases towards the flanks of the melting region.  This porosity
distribution sets the overal pattern of melt migration towards the
ridge axis.  Within c. 30~km either side of the spreading axis, melts
are focused to the ridge axis by buoyancy and lateral pressure
gradients, agreeing with predictions by \citet{Spiegelman1987} and
\citet{Yinting1991}.  At distances greater than 30~km, the mode of
melt focusing resembles the model by \citet{Sparks1991}, in which
magma rises vertically to the base of the lithosphere, and then uphill
along the top of the melting region to the ridge axis.

In numerical experiments initialised with a heterogeneous mantle,
magma flow localizes into high porosity channels.
Figure~\ref{fig:ridge} shows snapshots of the matrix composition
(figure~\ref{fig:ridge}b,e), porosity (figure \ref{fig:ridge}c,f) and
temperature ((figure~\ref{fig:ridge}d,g) from representative
experiments with $U_0 = 4 \cmyr$.  In these simulations, the
arrangement of channels and dynamics of focusing strongly depend on
the topology of mantle heterogeneity. In the random blob experiments
(figures~\ref{fig:ridge}b, c, d) some channels are narrow and
vertical, while others coincide
with smeared-out heterogeneities. Melt focusing to the ridge axis
is realised partly through channelized flow, and partly by flow along
the lithosphere--asthenosphere boundary.  In the smoothed noise
experiment (figures~\ref{fig:ridge}e, f, g) channels are wider and
focus magma to the ridge axis more efficiently.  Channels are
deflected around less fusible 
regions of the mantle, and tend to coincide with compositions that lie
between the average mantle and those moderately enriched in the more
fusible component. Although coexisting magma and mantle rock are in
local chemical and thermal equilibrium in all partially molten
regions, significant chemical disequilibrium can exist between magmas
within channels and magmas outside of channels.  This is because the
timescale for melt transport through channels is much shorter than the
timescale for diffusive equilibration of the channel and host mantle
\citep{Spiegelman1992}. 

In these examples, variation of porosity within the melting region is
a consequence of combined mechanical (compaction) and thermodynamic
(melting and freezing) effects.  \citet{Weatherley2012} showed that
the melting rate $\Gamma$ can be expressed as the sum of contributions
from mantle upwelling, reactive flow and thermal diffusion.  The
melting rate is such that in a homogenous mantle, mantle upwelling
accounts for over $95\%$ of the total melting rate. But in a heterogeneous
mantle where flow is channelized, upwelling can account for a minor fraction of the total melting
rate (as small as 20$\%$ in simulations by \citet{Weatherley2012}),
with the remained realised through reactive dissolution and thermal diffusion.
Thus porosity variations in a heterogeneous mantle are largely a
response to local thermodynamic effects.

\subsection{Duration and distance of melt migration}

\begin{figure*}
  \centering
  \includegraphics{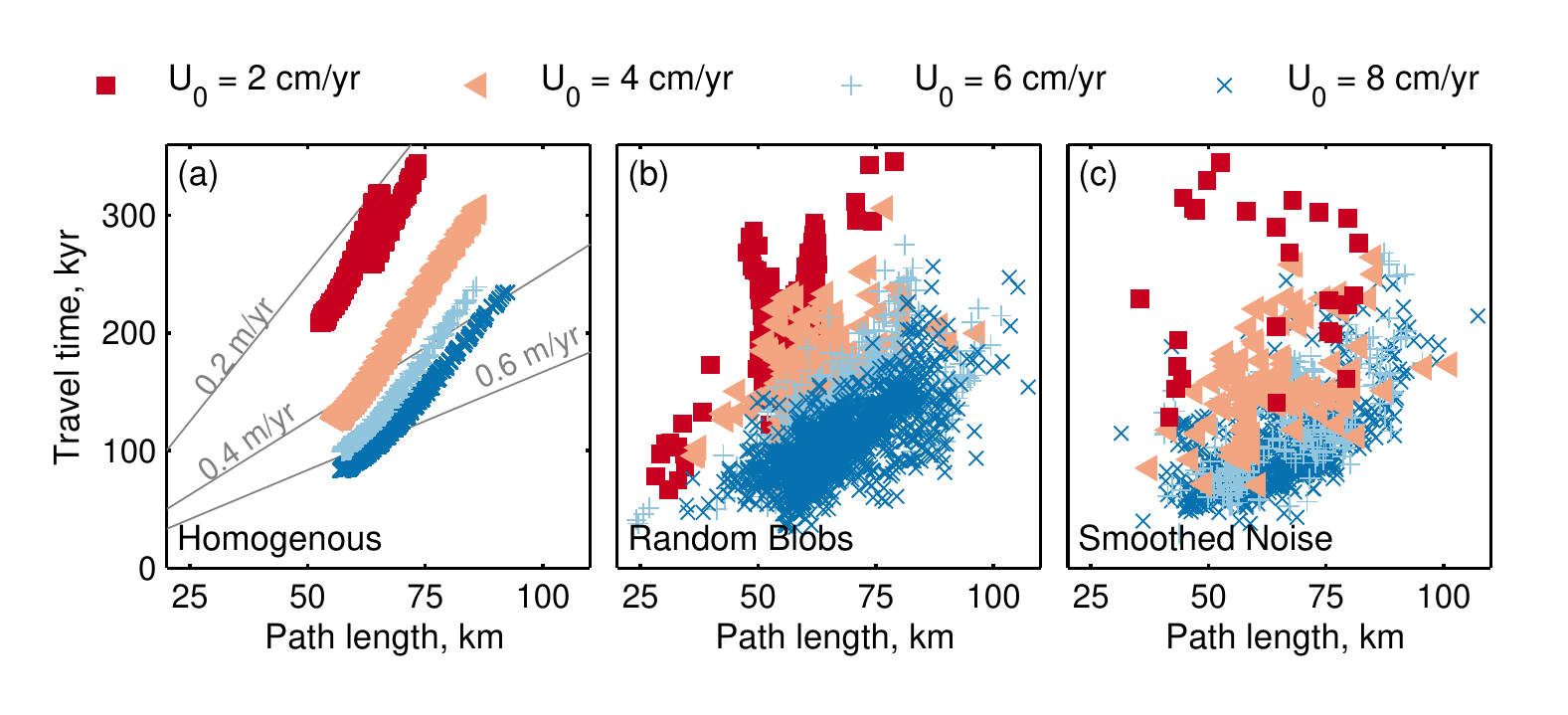}
  \caption{Travel time for individual particles plotted as a function
    of path length for half-spreading rates $U_0$ 2, 4, 6, and 8~cm/yr.
    Panel (a) shows data from simulations with a homogenous mantle.
    Grey lines correspond to the magma flow speeds indicated. The
    number of particles reaching the ridge axis $n$ in each
    simulation, arranged in order of increasing $U_0$ is 2168, 6663,
    11765 and 17821.  Panel (b)
    shows data from simulations where the heterogeneity is manifested
    as random blobs. For these simulations $n$ = 363, 805, 1380 and
    1828. Panel (c) plots data from simulations run with 
    smoothed noise heterogeneity. For these runs $n$ = 30, 103, 278
    and 575.}
  \label{fig:dist-time}
\end{figure*}

Figure \ref{fig:dist-time} plots the time taken for tracer particles
to travel from the base of the melting region to the ridge axis as a
function of the distance that they travel.  The duration of melt
migration for any tracer particle is also referred to as the travel
time, $\tau$, and the distance travelled by tracer particles $d$ is
interchangeably referred to as the path length.  These quantities
are defined as
\begin{equation}
  \tau = t_{\text{exit}} - t_{\text{release}},
  \label{eq:travel-time}
\end{equation}
\begin{equation}
  d = \int_{t, \text{release}}^{t, \text{exit}} | {\bf{v}}_f \left( t \right)| \ud t ,
  \label{eq:path-length}
\end{equation}
where $t_\text{release}$ is the time at which a particle is released
for the first time from the solid by melting and $t_\text{exit}$ is
the time at which the same particle exits the melting region at the
ridge axis.

Figure \ref{fig:dist-time}a shows results from experiments initialised
with a homogeneous mantle; the travel time increases approximately
linearly with the path length and is shorter when the spreading rate
is fast.  Path lengths range from 50--75 km when $\Uo = 2$~cm/yr to
55--95~km when $\Uo = 8$~cm/yr.  For each experiment the range of
travel times is approximately 150~kyr, and decreases from 200--350~kyr
when $\Uo = 2$~cm/yr to 90--240~kyr when $\Uo = 8$~cm/yr.  The arrays
of data are more densely populated with tracer particles representing
shorter path lengths and travel times since the upwelling rate is
fastest and porosity is highest directly beneath the ridge axis.  The
spreading rate dependence of travel time is discussed more fully in
section \ref{sec:sr}.

Figures \ref{fig:dist-time}b and \ref{fig:dist-time}c show results
from experiments run with the random blob and smoothed noise
heterogeneity models.  To leading order, these data show the same
trends as experiments with a homogenous mantle
(fig.~\ref{fig:dist-time}a); the travel time increases with path
length and decreases with spreading rate.  However, the data in
figures~\ref{fig:dist-time}b,c exhibit much scatter and the 
correlation between path length and travel time is weaker.  Compared
to results from the homogenous case, the data span a wider range of
travel times and path lengths.  For a half spreading rate of 2~cm/yr,
the range of travel times is approximately 50--350~kyr and path
lengths span 25--80~km; for $\Uo = 8$~cm/yr travel times span
25--300~kyr and path lengths range between 25~km and 110~km.  The
arrays of data in figures~\ref{fig:dist-time}b,c are more densely
populated for intermediate path lengths and travel times.  They show
that the duration of melt migration is shorter in the channelized flow
regime by around 10--20\%.  The sensitivity of these results to the
parameter regime for the model and discussion on their correspondence
to U--series disequilbria is presented in sections \ref{sec:scale} and
\ref{sec:implications}.

\subsection{Speed}

\begin{figure*}[h!]
  \centering
  \includegraphics{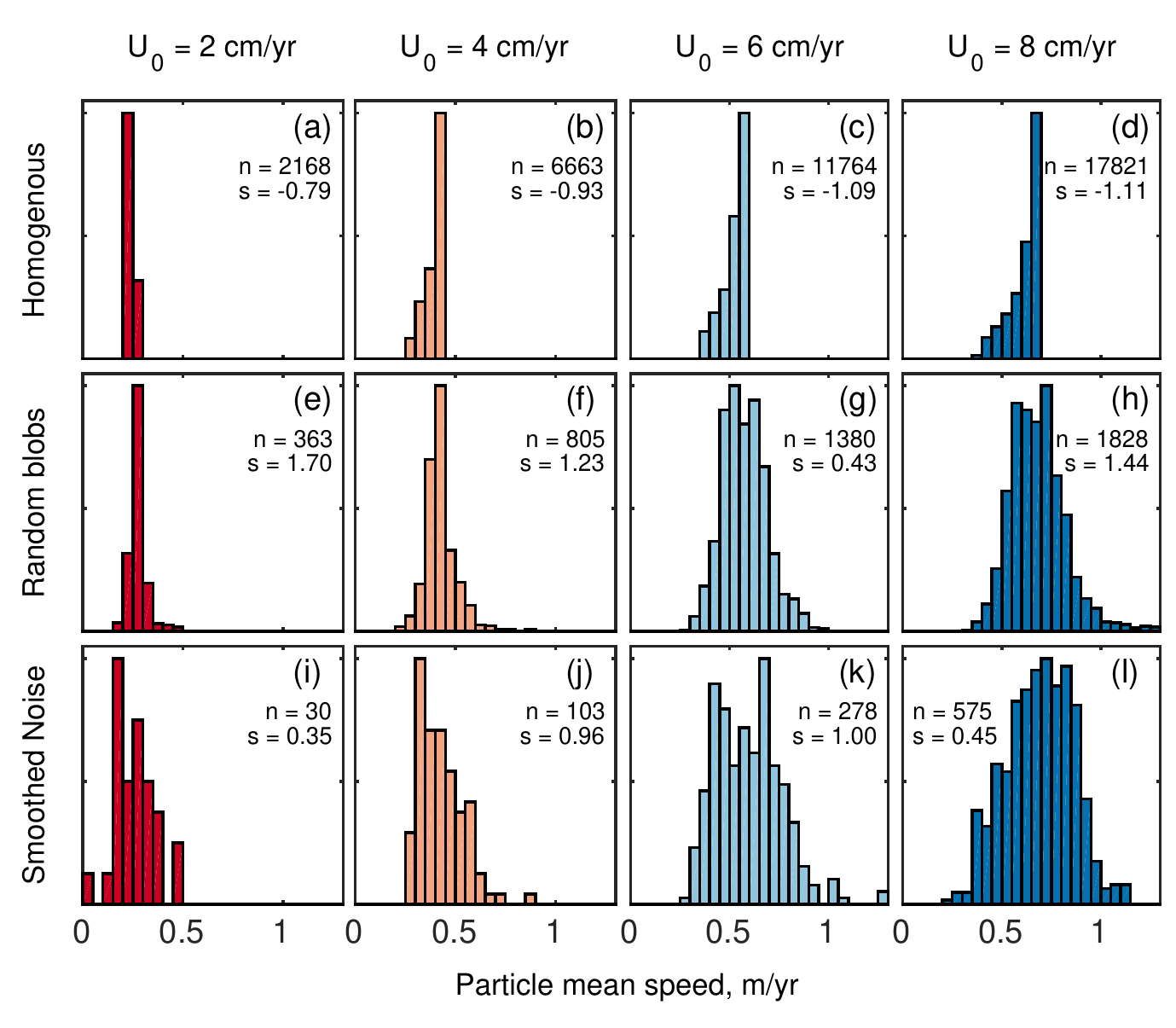}
  \caption{Normalized histograms showing the local mean speeds of
    particles.  Each column shows results from simulations with the
    same half-spreading rate and each row presents data from
    simulations with different styles of heterogeneity as the initial
    condition. Bin widths are 5~cm/yr, $n$ is the number of data
    points contained within the histogram, and $s$ is is an estimate of
    the population skewness.}
  \label{fig:hists}
\end{figure*}

Figure \ref{fig:hists} shows normalized histograms of the particle
mean speed for each of the numerical experiments presented in figure
\ref{fig:dist-time}.  The mean speed for particle $p$ is defined as
$\msp = \tau_p/d_p$, where $\tau_p$ and $d_p$ are the
particle-specific travel times and path lengths defined in
equations~\ref{eq:travel-time} and \ref{eq:path-length}. Figures
\ref{fig:hists}a--d show data from experiments run with a homogenous
mantle.  These histograms have a strong negative skew, revealing that
particles with the fastest speeds are the most abundant.  From
figure~\ref{fig:dist-time}a it is evident that the fastest particles
are generally associated with shorter travel times and path lengths.
The mode of the binned data increases with spreading rate from
0.20--0.25~m/yr at $\Uo = 2$~cm/yr to 0.65--0.70~m/yr when $\Uo =
8$~cm/yr.  The range of mean particle speeds also increases with
spreading rate from 10~cm/yr when $\Uo = 2$~cm/yr to 35~cm/yr when
$\Uo = 8$~cm/yr.

Results in figures \ref{fig:hists}e--h and \ref{fig:hists}i--l
correspond to experiments run with the random blob and smoothed noise
heterogeneity models.  These histograms have variable, positive skew
and show a factor of $\sim$2 increase on the range of mean particle
speeds predicted by experiments run without mantle heterogeneity.
Mean particle speeds range from 0--0.6~m/yr when $\Uo = 2$~cm/yr to
0.15--1.3~m/yr when $\Uo = 8$~cm/yr (figures~\ref{fig:hists}e--l).
Generally, histograms for the smoothed noise cases have a more
irregular structure than for those corresponding to experiments run
with the random blob heterogeneity model. The mode of the binned, mean
particle speeds is approximately independent of the presence or
topology of mantle heterogeneity.

\subsection{Effect of composition}

\begin{figure*}
  \centering
  \includegraphics{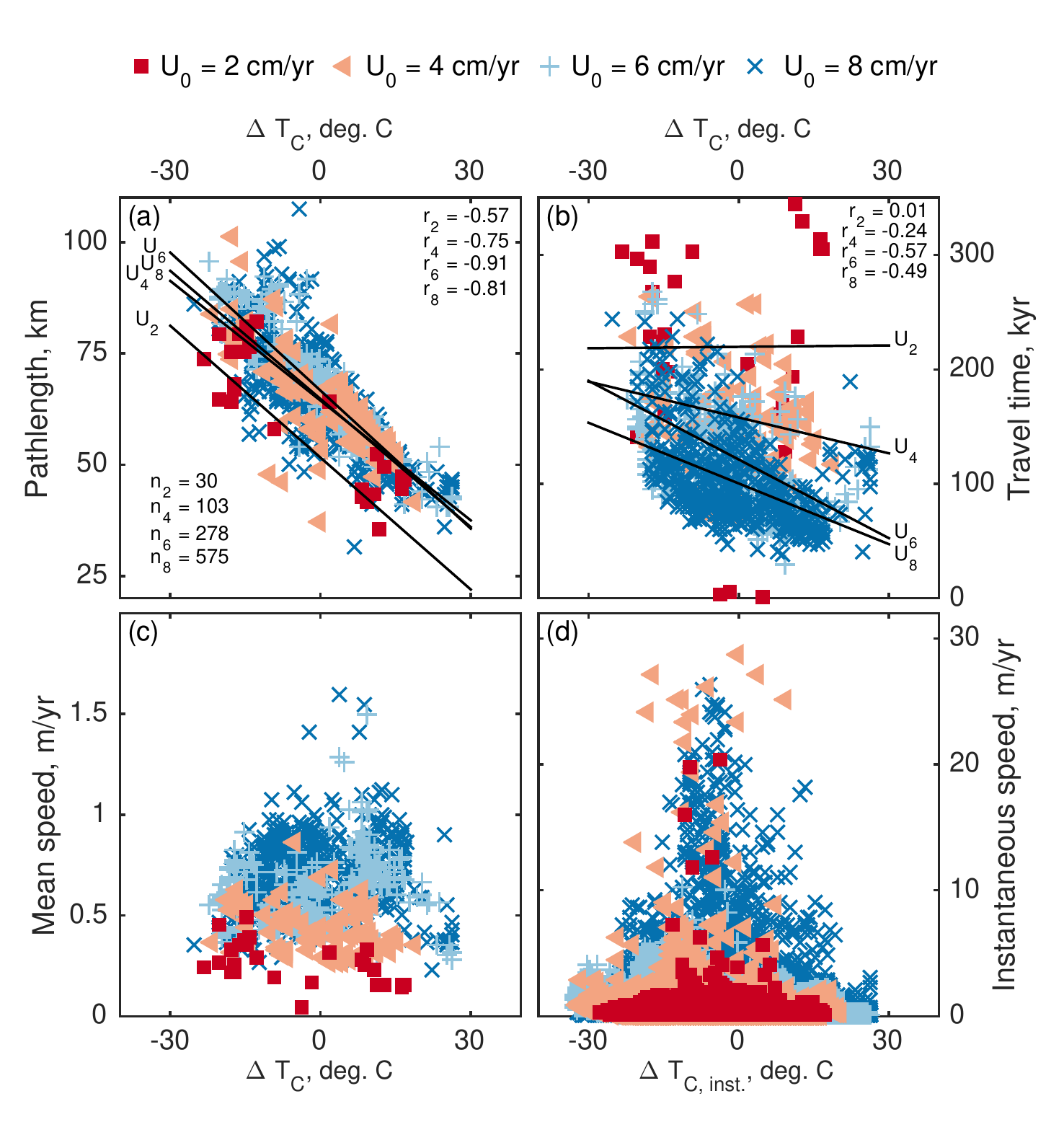}
  \caption{Effect of composition on the time scale and speed of melt
    migration for experiments run with the smoothed noise
    heterogeneity model.  (a) Variation of travel time with source
    composition $\Delta T$.  Black lines show least squares best
    fit lines for the different simulations and are labeled as U$_x$,
    where $x$ is the half spreading rate in cm/yr. The corresponding
    sample correlation $r$ for each of the best fit lines is given in
    the upper right of the figure, and $n_x$ is the number of data
    points from each simulation. (b) Path length as a function of
    source composition, with black lines as for panel (a).  Panel (c)
    plots mean particle speed as a function of source composition;
    panel (d) shows
    variation of the instantaneous speed with instantaneous
    composition, $\Delta T_{C, \mathrm{inst.}}$, expressed in terms of
    a perturbation in solidus temperature.  The sampling interval is
    approximately 1000~years.}
  \label{fig:SC}
\end{figure*}

Figure \ref{fig:SC} reveals the effect that the source fusibility at
the on set of melting, $\DTCO$, has on the duration and speed of melt
migration.  Data are taken from experiments run with the smoothed
noise heterogeneity model only.

Figure \ref{fig:SC}a shows that the path length correlates negatively
with $\DTCO$.  This negative correlation is a consequence of the fact
that more fusible rocks (more negative $\DTCO$) start to melt deeper.
The combined effects of the shape of the melting region, channelized
flow, and mantle heterogeneity mean that a range of path lengths can
be associated with any given composition.  Data are distributed
relatively evenly across source compositions between $-20\degC \leq
\DTC \leq 20\degC$ but are less abundant for more extreme values.
Figure~\ref{fig:SC}b shows that to leading order, the travel time and
source compositions are related by a weak to moderate negative
correlation for half spreading rates of 4, 6, and 8 cm/yr, although no
correlation is evident for $U_0 = 2$ cm/yr.  The leading order trend
arises from more fusible heterogeneities melting at deeper depth,
while the weak correlation implies that the connectivity of channels
and permeability structure does not vary systematically with source
composition.

Figure \ref{fig:SC}c plots the mean particle speed against
$\DTCO$. Although no obvious trend relates the two quantities, it can
be seen that mean speeds are slower for regions where the fusibility
perturbations are large ($|\DTCO|>20\degC$) and faster where
fusibility is perturbed to a lesser extent.  This result reinforces
the implication from figure \ref{fig:SC}b of non-systematic variations
in permeability structure and channel connectivity.

To complement these data, the speed and composition for each particle
were sampled at intervals of about 1000~years after the onset of
melting to give the instantaneous speed and instantaneous composition.  These
results, shown in figure~\ref{fig:SC}d, exhibit negligible correlation
between speed and composition along trajectories of tracer particles.
However, the greatest range and fastest speeds are observed in regions
where mantle heterogeneity perturbs the solidus temperature between
$-15\degC$ and $+10\degC$.  In these regions, instantaneous speeds
are, in some instances, as fast as 30~m/yr, but are often as low as a
few decimetres per year.  For more extreme compositions, the range and
magnitude of instantaneous speeds is much reduced.  In these regions
the maximum magma speeds are around 2~m/yr, but the vast majority of
tracers passing through more highly enriched and depleted regions have
speeds of a few centimetres to a few tens of centimetres per year.

\section{Discussion}
\label{sec:discussion}

\subsection{Effects of plate spreading rate}
\label{sec:sr}

\begin{figure*}
  \begin{center}
    \includegraphics{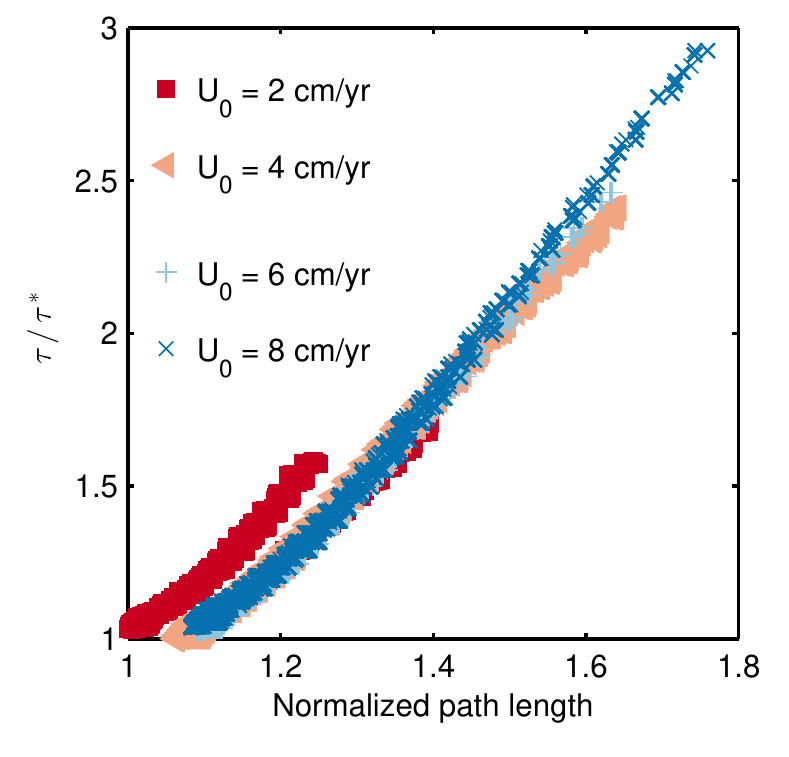}
    \caption{Scaled travel time versus nondimensional distance for the
      experiments conducted with no mantle heterogeneity.  Travel time
      $\tau$ is scaled by $\tau^* = (3/2 U_0)^{1-1/n}/l$, where $l$ is
      a characteristic length scale.  }
    \label{fig:scaling}
  \end{center}
\end{figure*}

Two leading-order trends emerge in the results: that shorter
melt-transport path lengths and faster spreading rates are both
associated with shorter durations of melt migration.  These trends
arise because plate spreading confines melting to a roughly
triangular--shaped region of non--constant porosity.  Analysis of the
governing equations in 1D by \citet{Hewitt2010} reveals that the magma
speed $w$ scales with $W^{1 - 1/n}$, where $W$ is the upwelling rate
of the matrix and $n$ is the permeability exponent \citep{Bear1972,
  Turcotte2002, Costa2006}.  For mantle convection patterns resembling
corner flow, the half spreading rate defines a reference upwelling
rate $W_0 = 3/2U_0$ \citep{Batchelor1967}, thus $w \propto
U_0^{1-1/n}$.  Figure \ref{fig:scaling} shows that this scaling
accounts for the spreading rate dependance of the travel time in a
homogenous mantle.  In the experiment with $U_0 = 2~\cmyr$, tracers
associated with short path lengths ($d<65$~km) have anomalously slow
speeds.  This is a consequence of the slow upwelling causing smaller
melting rates, lower porosity, lower permeability, and hence slower
segregation. 

\subsection{Effects of heterogeneity}

\begin{figure*}
  \begin{centering}
    \includegraphics{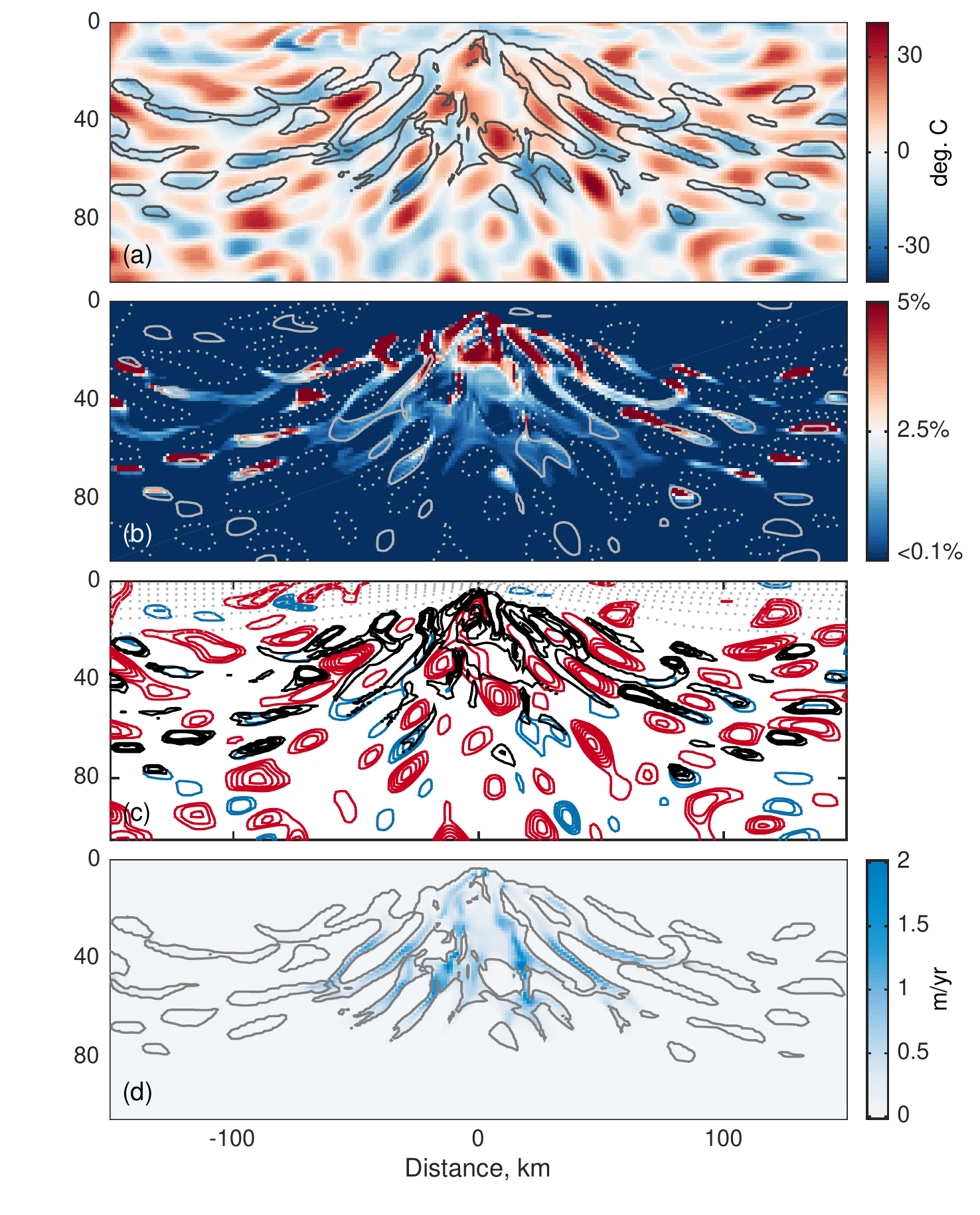}
    \caption{Covariation of porosity and composition from a
      representative experiment initialized with the smoothed noise
      heterogeneity model and $U_0 = 6$ cm/yr.  (a) Colours show
      composition in terms of a perturbation to the solidus
      temperature; grey lines contour $\phi=1\%$. (b) Colour shows
      porosity; grey solid lines contour $\DTC = -15\degC$; grey
      dotted lines contour $\DTC = +10\degC$. (c) Covariation of large
      porosities and regions of enriched and depleted mantle.  Black
      lines contour $\phi = 1\%$, blue lines contour
      $\DTC\leq-15\degC$ in $5\degC$ intervals, and red lines
      contour $\DTC \geq +10\degC$ in $5\degC$ intervals.  (d) Magma
      speed $|\vf|$ with grey lines showing the 0.5\% porosity contour.
    }
    \label{fig:pc}
  \end{centering}
\end{figure*}

Compared to plate spreading, mantle heterogeneity has a more
complicated effect on melt migration.  While the fact that less
fusible mantle melts at shallower depths readily explains the negative
correlation between travel time and source composition, the
relationship between speed and composition is more difficult to
understand.  To help explore why faster speeds are associated with
source and regions where perturbations to the fusibility are
relatively minor ($|\DTC|<20\degC$, figures~\ref{fig:SC}c, d),
figure~\ref{fig:pc} compares the spatial covariation of mantle
heterogeneity with porosity for a representative experiment run with
$U_0 = 6~\cmyr$ and the smoothed--noise heterogeneity model. In panels
a and c, black lines contour porosities $>1\%$ to roughly show the
location of channels and magma ponds.  In this experiment, channels
are confined, more or less, to the region within 100~km of the
spreading axis and depths $<60$~km.  The channels accommodate the
fastest magma speeds (figure \ref{fig:pc}d) and broadly coincide with
mantle of intermediate fusibility ($-15$~K $\leq \DTC \leq+10$~K); in
regions of more extreme fusibility the flow speeds are slower.  This
correlation is also evident in figure~\ref{fig:SC}d.

This relationship between composition and magma speed is a consequence
of the energetics of melting.  The imposition of local thermodynamic
equilibrium pins the temperature of partially molten regions to the
solidus temperature.  Thus heat diffuses away from locally depleted
regions into the surrounding more fusible and cooler mantle.  In the
most depleted regions ($\DTC \geq 10$K), therefore, melting is
suppressed, porosity is destroyed by compaction and the resulting
small permeability accommodates slow instantaneous magma speeds and
acts to decrease the mean speeds of magmas originating in these
regions.  Flow speeds in the most enriched regions ($\DTC \leq -15$K)
are slow for a different reason: enriched heterogeneities start to
melt before the surrounding mantle and are surrounded by a halo of
impermeable mantle rock.  This prevents magma from flowing into the
channel network; within the heterogeneity the magma speed $|\vf|$
approximates the mantle speed $|\vm|$, which results in slow
instantaneous and mean speeds.  It is interesting to note that,
since the average flow speed for each particle is, in part, a
response to local thermodynamic conditions, different
heterogeneity configurations could result in speed-composition
distributions different to those shown in figures \ref{fig:pc}c and
d. 

\subsection{Effects of model parameters}
\label{sec:scale}

An important point about the results above is that they are sensitive
to the model parameters, and particularly those that directly affect
the magma velocity.  \ref{sec:scales} presents a nondimensionalization
of equation \ref{eq:time-vf} for $ {\bf{v}}_f$ and shows that the
nondimensionalizing factor is $\mathcal{R} = k_0 \drho g / \mu$, where
$k_0$ is the reference permeability, $\drho$ is the density difference
between the matrix and magma, $g$ is the acceleration due to gravity
and $\mu$ is the viscosity of the magma. Of these, $g$ is accurately
known, possible values for $\drho$ vary to within a few percent of the
value given in table \ref{tab:ridge-p-vals} (500~kg~m$^3$),
measurements of $\mu$ span one order of magnitude (1--10~Pa$\cdot$s,
\citet{Dingwell1995}) and estimates of $k_0$ lie in the range
$10^{-9}$~m$^2$ to $10^{-4}$~m$^2$ \citep{Faul1997, Wark1998,
  Faul2001, Wark2003, Connolly2009}.  Uncertainty in these parameters
means that the range of plausible values for $\mathcal{R}$ spans some
6 orders of magnitude.  For comparison, a permeability constant of
$k_0 = 10^{-7}$m$^2$ was selected for the model runs above, and the
corresponding value of $\mathcal{R}$ lies approximately in the centre
of the possible range.

\setcounter{table}{0} \renewcommand{\thetable}{A\arabic{table}}
\begin{table}[h!]
  \begin{center}
    \caption{List of parameters and preferred values.}
    \begin{tabular}{ l p{6.75cm} l }
      \hline
      Symbol & Meaning&Value\\
      \hline
      $\kappa$&Thermal diffusivity&$1\times 10^{-6}$ m/s$^2$\\
      $L$& Latent heat of fusion&$5.5\times10^5$J/kg\\
      $M_S$&Slope $\partial T/\partial C$ of the solidus& 200 K\\
      $M_L$&Slope $\partial T/\partial C$ of the liquidus& 200 K\\
      $\mathcal{T}$& Potential temperature&1696 K\\
      $T_0$& Solidus temperature for reference mantle at 0 kbar& 1605 K\\
      $C_0$&Reference mantle composition&0.5\\
      $\Delta C$&Difference in composition between rock and magma in
      thermodynamic equilibrium&0.1\\
      $\Delta T$ & Temperature scale & $M\Delta C$\\
      $\Delta \rho$&Matrix--melt density difference& 500 kg/m$^3$\\
      $\alpha$&Coefficient of thermal expansion&$3\times10^{-5}$ K$^{-1}$\\
      $c_P$&Specific heat capacity&1200 J/kg/K\\
      $\gamma^{-1}$& Clapeyron slope& 60 K/GPa\\
      $\mathcal{D}$&Chemical diffusivity&$1\times10^{-8}$ m/s$^2$\\
      $g$&Acceleration due to gravity&9.8 m/s$^2$\\
      $k_0$& Reference permeability& $1\times 10^{-7}$ m$^2$\\
      $\eta_{\mathrm{max}}$&Cut-off shear viscosity,
      maximum&$10^{25}$Pa$\cdot$s\\
      $\Delta x$ & Grid spacing in $x$-direction & 1.25 km\\
      $\Delta z$ & Grid spacing in $z$-direction & 1.25 km\\
      \hline
    \end{tabular}
    \label{tab:ridge-p-vals}
  \end{center}
\end{table}

\begin{figure*}
  \includegraphics{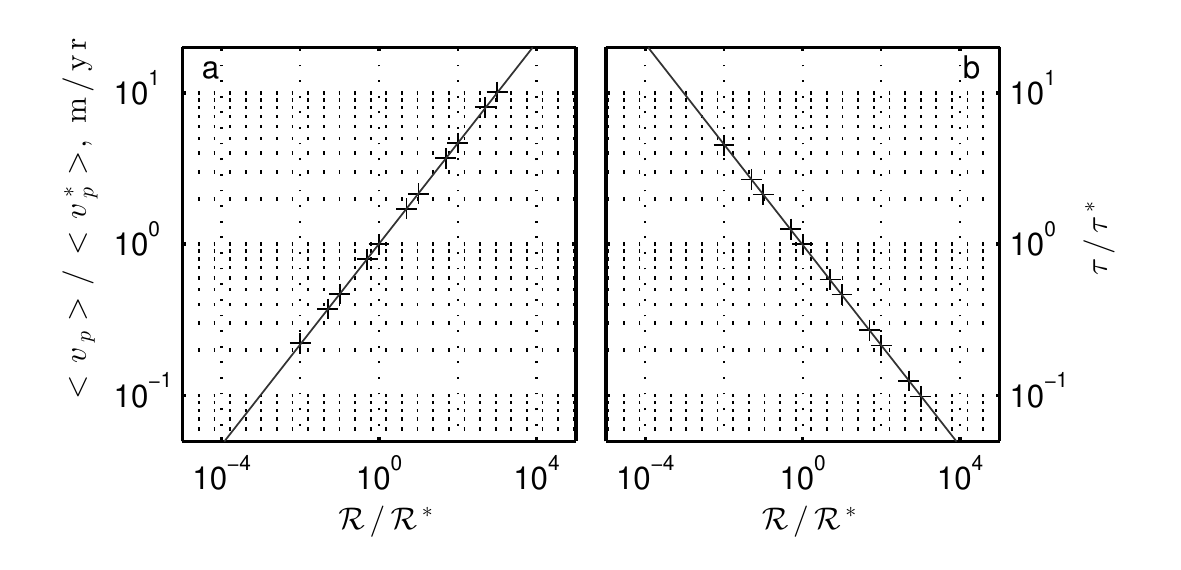}
  \caption{The effect of $\mathcal{R} = k_0 \drho g / \mu$ on melt
    migration computed using the upwelling column model outlined in
    \citet{Weatherley2012}.  The upwelling rate is 6 cm/yr and the
    grid spacing is 1 km.  Other parameter values
    are given in table \ref{tab:ridge-p-vals}. (a) Effect of
    $\mathcal{R}$ on the mean particle speed $\langle v_p \rangle$.
    $\mathcal{R}^*$ and $\langle v_p \rangle^*$ indicate normalizing
    values computed from an experiment conducted with the parameter
    regime in table \ref{tab:ridge-p-vals}. (b) Effect of
    $\mathcal{R}$ on the duration of melt migration $\tau$, where
    $\tau^*$ is associated with the $\mathcal{R}^*$ parameter regime.}
  \label{fig:R}
\end{figure*}

To investigate how the mean particle speed $\langle v_p \rangle$ and
travel time $\tau$ vary with $\mathcal{R}$, the model was reconfigured
to consider the dynamically simpler case of a column of upwelling,
chemically homogenous mantle rock. Figure \ref{fig:R} plots the
predicted $\langle v_p \rangle$ and $\tau$ for values of $\mathcal{R}$
that span 5 orders of magnitude.  The data are normalized by results
from the model run with the parameter regime specified in table
\ref{tab:ridge-p-vals}. The figure shows that $\langle v_p \rangle$
scales with $\mathcal{R}^{1/n}$ (figure \ref{fig:R}a), where $n$ is
the permeability exponent and that $\tau$ varies with
$\mathcal{R}^{-1/n}$.  This scaling agrees with analysis of the
governing equations in one dimension by \citet{Hewitt2010}; it
indicates that an change in $\mathcal{R}$ by a factor of 10$^3$ will
yield an order of magnitude change in $\langle v_p \rangle$ and
$\tau$.  Provided that the scaling holds for models with a
heterogeneous mantle, setting $\mathcal{R} = 10^3 \mathcal{R}^*$ would
reduce the average travel times to order $10^4$ years and
increase mean flow speeds to order 10~m/yr.

\subsection{Limitations and assumptions}
\label{sec:limitations}

The results above are also subject to a set of limitations arising
from the assumptions that underpin the models and theory.  An
important assumption is that melt migration is accommodated entirely
by porous flow \citep{McKenzie1984, Fowler1985, Ribe1985,
  Bercovici2001}.  Whilst a mass of geochemical and geophysical
observations require the mantle to be porous and melt to migrate by
porous flow \citep[e.g.][]{Johnson1990, Johnson1992, MELT1998,
  Sims1999}, several studies also suggest that melt can flow through
cracks \citep{Sleep1988, Fowler1996, Ito2002, Maaloe2003}.  This mode
of magma flow is less well studied than reactive channelisation, thus
uncertainty surrounds how widely it applies to the sub-ridge melting
region.  Nevertheless, magma flow through open cracks can accommodate
much faster flow speeds than melt migration through high porosity
channels; hence this mode of localisation lends itself well to
arguments for preservation of deep--origin $\Pb$ disequilibria in
zero-age MORB \citep{Rubin2005}. Future models should seek to include
a brittle fracture criterion in order to assess the implications of
fracture for melt extraction.

An additional limitation lies in the two-dimensional description of
the dynamics at mid-ocean ridges. This imposes an infinite,
out-of-plane length on all heterogeneities and melt migration features
in the model and restricts mass and heat transport to be in the plane
of the solution. \citet{Katz2012} note that this latter restriction
could reduce the efficiency of channel formation.  For instance, if a
channel has a cylindrical geometry in three dimensions, mass and heat
can be supplied from around its circumference.  Radial heat and mass
transfer could potentially reinforce the channel-forming instability,
resulting in higher porosities and faster magma flow speeds.  Three
dimensional consideration of melt migration is also important in the
vicinity of ridge offsets.  Previous studies show that hot mantle
beneath ridge segment ends is cooled by the adjacent, older and cooler
lithosphere \citep{Magde1997, Weatherley2010, Herbert2011}.  This
cooling effect is likely to suppress porosity generation, result in
lower magma flow speeds, and focus melt towards segment centres.
Exactly how the presence of mantle heterogeneity might modify melting
and melt transport close to segment ends is not yet understood, but
could have particular relevance to the interpretation of bull's eye
gravity anomaly patterns along highly segmented ridges, such as the
Mid-Atlantic Ridge \citep{Kuo1988, Lin1990, Detrick1995, Magde1997}.

The petrological model is the most significant simplification in our
theoretical framework for magma flow.  Motivated by experimental
studies of mantle rocks, the model captures variation of the mantle
solidus with composition and pressure, and takes an energetically
consistent approach to melting by decompression and reaction.  These
properties approximately align the model with natural system.  For
simplicity, however, the model system comprises two phases and two
thermodynamic 
components.  Whilst this precludes consideration of additional
important petrological features of the sub-ridge mantle, such as
variations in modal mineralogy, volatiles, phase specific enthalpies
of fusion, and reactive precipitation of pyroxene, it lends itself
well to elucidating the dynamical consequences of mantle heterogeneity
for melt migration.  A particularly important result of the
counterpart models by \citet{Weatherley2012} and \citet{Katz2012} is
that the arrangement and existence of melt migration pathways is a
response to local thermodynamic effects. This finding is expected to
hold both on smaller length scales and in simulations with an expanded
petrological model; thus the current model provides a reasonable,
first order description of the large scale dynamical behaviour.
However, a more detailed consideration of the petrology may offer new
insights to trace element redistribution and generation of U-series
disequilibria in the melting region.

Similar to earlier models by \citet{Katz2010}, magma is extracted at
the ridge axis.  The internal boundary that facilitates melt
extraction is carefully tuned so that on time scales that are long
compared to a time step, the rate of magma production approximately
matches the combined rates of melt extraction at the ridge, freezing
at the base of the lithosphere and exit of magma through the side
walls of the domain.  One consequence of this internal boundary is a
slight build-up of magma in the top 20 km of the mantle beneath the
ridge axis.  In this region, the dynamic pressure gradients are such
that the magma velocity is small, despite the large porosity. Regions
of high porosity are also predicted to occur in isolated patches in
the flanks of the melting region.  Since the permeability increases as
a power-law of porosity, magma within these regions can convect at
speeds of hundreds of metres per year, drastically reducing the size
of the time-step in the simulations.  To avoid this, the permeability
is capped at a value corresponding to a porosity of $10\%$.  For the
most part, the permeability cap has negligible effect on the speed of
melt migration within channels.  It also has negligible effect on
magma speeds in the region of slightly higher porosity immediately
beneath the ridge axis, owing to the setup of the internal boundary
condition for melt extraction.

An additional limitation of the model is the grid resolution.  In the
simulations above, the grid resolution is 1.25~km, which limits the
smallest length scale of heterogeneity and width of melt localization
features. A key result of the counterpart models by
\citet{Weatherley2012} and \citet{Katz2012} is that the arrangement
and existence of melt migration pathways is response to local
thermodynamic effects.  Therefore, it is reasonable to expect a
similar basic model behaviour in simulations run with much finer grid
resolution.  Alongside offering opportunities to explore the coupling
between heterogeneity on sub-kilometer scales and magma flow speeds,
finer resolution models offer greater possibilities for investigating
how the number density of fusible and refractory heterogeneities
influence the large scale pattern of melt migration.  Many studies
of ophiolites and peridotite massifs demonstrate
that mineralogical heterogeneity can be present on scales upwards of a
few centimetres \citep[e.g.][]{Boudier1981, Gregory1984, Reisberg1991,
  Blichert1999, LeRoux2007}. 
At these small scales, physical attributes of the mantle such as grain
size variations and minor shear zones are likely to have consequences
for melt localization, in addition to those from composition
heterogeneity.  Our understanding of how these different variables
influence the dynamics of magma flow and hence the time scale of melt
migration will benefit from future studies that combine lithological
and structural mapping of mantle peridotites on a range of scales with
higher resolution numerical models with more advanced rheological laws
than that used here.

\subsection{Geochemical implications}
\label{sec:implications}

A major scientific challenge at present is to understand the chemistry
of oceanic basalts in terms of source composition and processes that
affect magmas between the point of origin and eruption.  A key result
from the models above is that under conditions of strong mantle
heterogeneity, melts migrate to the ridge axis on a broad range of
time scales.  We postulate that {\textit{(i)}} this distribution in
travel times maps on to significant natural variability in the
U--series disequilibria observed along MORs, and {\textit{(ii)}} that
the results offer an explanation for the decoupling that is often
observed between U-series disequilibria and other geochemical data,
even in settings where there is evidence for mineralogical source
heterogeneity \citep[e.g.][]{Kokfelt2003, Bourdon2005, Stracke2003,
  Prytulak2009, Russo2009, Koornneef2012}. To further understand the
implications of these results, future studies should aim to constrain
how properties of mantle heterogeneity and the melt transport network
are linked.  Possible avenues to investigate include how the number
density, fusibility and connectedness of heterogeneity relate to the
topology and tortuosity of melt channels, whether channel tortuosity
on smaller scales than that examined here has a significant effect on
melt transport time, and how heterogeneity and magma flow together
affect the rate of magma supply to ridges, and by extension their
morphology.

Within the context of the model results, it is also important to
consider other processes that can have notable implications for the
geochemistry of oceanic basalts.  Among the most significant is
mixing.  Analyses of whole rock and melt inclusion chemistry by
\citet{Maclennan2003} and \citet{Maclennan2008} demonstrate that
erupted magmas are mixtures of many small batches of melt, with up to
30 batches of melt represented in a single hand
specimen. Additionally, \citet{Rudge2013} showed that the mixing
process is nonuniform with depth, with the deepest melts appearing to
be more homogenized than shallower melts.  Interestingly, they also
found that compositional arrays produced during mixing do not
necessarily point towards the isotopic compositions of the source
materials.  A second set of processes important to the the
preservation or homogenisation of primary melt compositions are those
that occur in the crust and uppermost mantle.  \citet{Rubin2007}
showed that high melt supply rates (typically associated with fast
spreading ridges) promote larger and shallower accumulations of cooler
magma, longer residence times in sub-ridge magma reservoirs, and
consequently greater extents of homogenization and differentiation.
In contrast, lower melt supply rates promote episodic magma
accumulation in hotter, more poorly connected magma reservoirs, and
erupted products that show high degrees of chemical variability and
lower degrees of differentiation.  In the shallow crust,
crystallisation processes and melt--rock reaction can obfuscate the
relationship between primary melt chemistry and that of erupted magmas
\citep[e.g.][]{VanOrman2006, Maclennan2008}. However,
\citet{Rubin2009} note that few sections of the MOR system are sampled
at high enough resolution to quantify how these processes, and by
extension, results of numerical models, relate to MORB compositions
and MOR structure.

Despite these additional complexities and limitations, the results
above also have several general implications for our understanding of
how theories of magma flow relate to observations of U-series
disequilibria in MORB.  Under the parameter regime outlined in table
\ref{tab:ridge-p-vals}, the results indicate that magma flow in a
heterogeneous mantle is most commonly fast enough to preserve $\Th$
excesses generated deep in the melting region.  Some particles travel
slowly enough for $\Th$ initial excesses to decay to zero.  However,
significant excesses for these particles could be generated by
ingrowth, or alternatively, their lack of $\Th$ disequilibrium could
be obscured by mixing.  For the same simulations, magma flow too slow
to preserve deep-origin $\Ra$ disequilibria, yet
section~\ref{sec:scale} suggests that under less moderate parameter
regimes, the models will predict flow speeds sufficiently fast to
enable attribution of some $\Ra$ disequilibria to a deep source.
Under even the most favourable parameter regime, the current
theoretical framework for porous melt migration is incapable of
preserving $\Pb$ disequilibria generated within the melting region.

To clarify this situation, and to address whether current theory can
account for $\Ra$ disequilibria, future studies must consider the full
range of disequilibrium--generating processes.  These include
chromatographic fractionation, ingrowth, degassing and diffusive
partitioning between melt and cumulate rocks.  Furthermore, the models
must also be able to account for covariation of different
parent-daughter nuclide pairs.  Models by \citet{Jull2002} suggest
that magmas from different portions of the melting region must mix to
produce the observed signature, and \citet{Elliott2003} demonstrate
that so called `full reactive transport models', which couple
equations for element concentration to dynamical models similar to
those described above, also predict an inverse relationship between
$\Ra$ and $\Th$ disequilibria, but fail to match the observed
amplitudes.  However, future modelling studies that consider U-series
disequilibria in the context of mantle heterogeneity and the physics
of melt migration may yield a more conclusive view on the presence of
short-lived U-series disequilibria in oceanic basalts.

\section{Conclusions}
\label{sec:conc}

Results from energetically consistent models of magma flow show that
plate spreading and mantle heterogeneity have important consequences
for the time scale of melt migration beneath ridges. Heterogeneity
perturbs the thermal structure of the mantle that is established by
plate spreading, and causes magma flow to localize into high porosity
channels.  Magma flows relatively rapidly through channels directly
beneath the ridge axis where the upwelling rate of the porous mantle
is greatest, and more slowly through channels on the flanks of the
melting region. Flow speeds also increase with spreading rate,
consistent with simpler models. The main effect of mantle
heterogeneity on the time scale of magma flow is realised through the
formation of high porosity channels.  However, magma flow is generally
slower in more enriched and more depleted regions, and faster in
regions of close to average fusibility.

Our findings suggest that the mean speed of channelized flow in a
heterogeneous mantle is approximately $20\%$ faster than diffuse flow
in a homogenous mantle.  For the parameter regime and spreading rates
considered, individual packets of magma in a single simulation migrate
to the ridge axis on time scales as short as 25~kyr and as long as
350~kyr, with mean speeds between 0.2 and 1.3~m/yr. Scaling analysis
suggests that for plausible parameter regimes that capture a more
permeable mantle, the duration of melt migration might be reduced by a
factor of 10.  These speeds are sufficient for mineral--melt
partitioning deep in the melting region to contribute to $\Th$ and, in
some circumstances even $\Ra$ disequilibria observed in MORB.
However, the models' inability to explain observations of $\Pb$
deficits in MORB suggests that current theories of porous melt
migration do not comprehensively describe the physics of magma flow.  

An important result is that mantle heterogeneity induces significant
natural variability in U--series disequilibria preserved in MORB, and
thus the models offer one explanation as to why observations are
often decoupled from other geochemical signals and properties of ridge
axes, even in situations where there is good evidence for
mineralogical heterogeneity in the mantle.  

\paragraph{Acknowledgements}
We wish to thank J. Maclennan and I. Hewitt for
interesting discussions, and K. Rubin and A. Stracke for
useful reviews; all of which helped to improve this manuscript.  Oxford
Supercomputing Centre, a former incarnation of the  Advanced Research
Computing facility at the University of Oxford is acknowledged for
providing compute time, HPC resources and technical support. SW thanks
the Danish Council for Independent Research for generous funding
whilst writing this manuscript.  RK thanks the Leverhulme Trust for support

\appendix
\section{Parameter Values}

The parameter values used for the numerical experiments above are
listed in table \ref{tab:ridge-p-vals}, below.

\section{Nondimensionalization}
\label{sec:scales}

To nondimensionalize equation \ref{eq:time-vf} the following scales
are employed:
\begin{equation}
  \left( {\bf{v}}_f^{'},  {\bf{v}}_m^{'} \right) = 
  w_0
  \left( {\bf{v}}_f, {\bf{v}_m} \right),
\end{equation}
\begin{equation}
  K^{'} = k_0 K,
\end{equation}
\begin{equation}
  \nabla^{'} = \frac{1}{H} \nabla,
\end{equation}
\begin{equation}
P_d = \Delta \rho g H P_d,
\end{equation}
where nondimensional variables are indicated with a prime, $H$ is the
height of the domain and $k_0$ is the reference permeability.
Application of these scales reveals that equation \ref{eq:time-vf} is
nondimensionalized by the factor $\mathcal{R} = k_0 \Delta \rho g /
\mu$.

\bibliographystyle{elsarticle-harv} \bibliography{references.bib}

\end{document}